# 4πβ (LS)-γ (HPGe) Digital Coincidence System Based on Synchronous High-Speed Multichannel Data Acquisition *

Jifeng Chen(陈吉锋)[1]    Kezhu Song(宋克柱)[1]    Juncheng Liang(梁珺成)[2]    Jiacheng Liu(柳加成)[3]

[1] State Key Laboratory of Particle Detection and Electronics, USTC, Hefei, 230026, China
[2] National Metrology Institute, Beijing 100029, China
[3] Ministry of Environmental Protection of Nuclear and Radiation Safety Center, Beijing 100082, China

**Abstract**: A dedicated 4πβ (LS) −γ (HPGe) digital coincidence system has been developed in this work, which includes five acquisition channels. Three analog-to-digital converter (ADC) acquisition channels with an acquisition resolution of 8 bits and acquisition rate of 1GSPS (sample per second) are utilized to collect the signals from three Photo multiplier tubes (PMTs) which are adopted to detect β decay, and two acquisition channels with an acquisition resolution of 16 bits and acquisition rate of 50MSPS are utilized to collect the signals from high-purity germanium (HPGe) which are adopted to detect γ decay. In order to increase the accuracy of the coincidence system, all the five acquisition channels are synchronous within 500ps. The data collected by the five acquisition channels will be transmitted to the host PC through PCI bus and saved as a file. Off-line software is applied for the 4πβ (LS)-γ (HPGe) coincidence and data analysis as needed in practical application. With all the above preconditions, the flexibility of the system is increased, and the structure and application of the system are simplified. According to the test, the highest coincidence rate of the system is 20K per second, which is sufficient for most applications. This paper mainly introduces the design of the hardware, the synchronization method and the test result of this system.

## 1 Introduction

The 4πβ-γ coincidence counting method has been a major technique for radionuclide standardization for decades. As a direct activity measurement method, it can determinate the activity without any quench indicating parameters, and the counting efficiency without reference standard[1]. Traditional 4πβ-γ coincidence system is composed of numerous electronic modules such as counting module, pulse shaping module, pulse-height analyzer and dead-time processing module. As shown in the Fig.1, the traditional 4πβ (LS) −γ (HPGe) consists MAC3, coincidence, and Counter module, it's very inconvenient. So far, new techniques and digital coincidence systems are applied for 4πβ-γ coincidence in various applications. The ADC and Field Programmable Gate Array (FPGA) were applied widely in the coincidence measurement to improve the performance of the coincidence system[1, 2, 3]. In the same time, the most of the dedicated modules like the MAC3 and MTR2 have been implemented by the logic of FPGA[4, 5, 6]. ADCs are adopted by these digital coincidence systems to sample the signal of 4πβ-γ system and FPGA fulfils the function of coincidence.

Normal digital 4πβ-γ coincidence systems employ the commercial acquisition card to sample the radionuclide decay signal. Most of these cards sample signal at a speed of 25MSPS max which are only fit to the requirement of 4πβ-γ coincidence with a detector configuration of a sandwich type plastic scintillator (PS) or a type NaI (Tl) scintillation detector[4]. To achieve the primary activity measurements based on LS such as 4πβ(LS) and triple to double coincidence ratio (TDCR), commercial acquisition cards with a higher acquisition speed like 125MSPS or 200MSPS is used[1,2]. But the pulse width of the liquid scintillation sometimes is 4ns[7], as the sampling interval of these coincidence system is 5ns, so these pulses probably can't be recorded. Although algorithm can be used to compensate the measurement error, deviation sometimes is still introduced. On the other hand, shape analysis can't play an efficiency role by using the data recorded by the acquisition channels mentioned above.

Manuscript submitted June 17, 2015

* Supported by National Ministry of Environmental Protection
  E-mail: skz@ustc.edu.cn

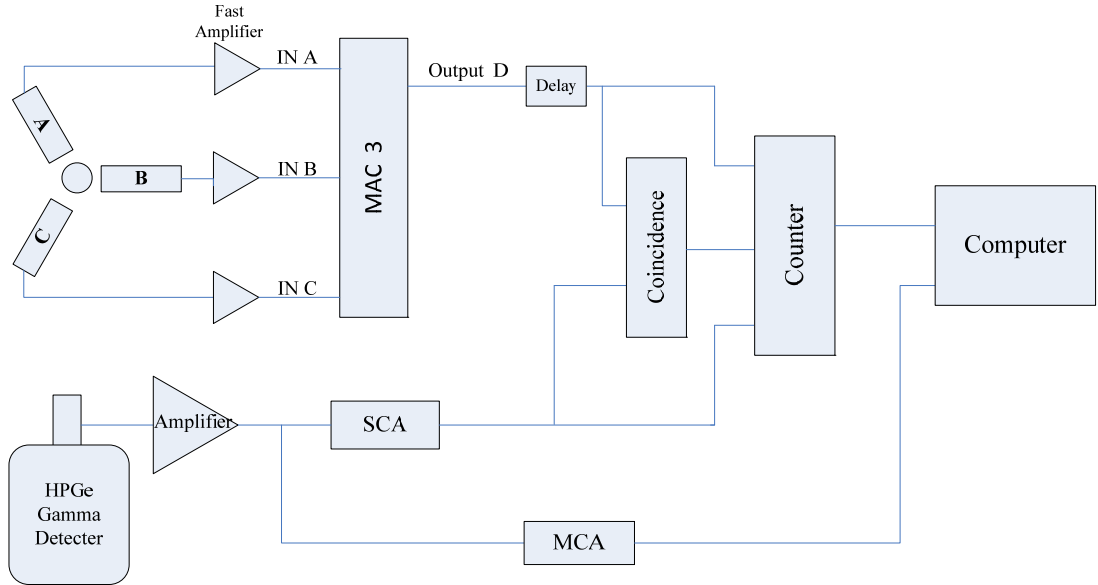

Fig. 1. Traditional 4πβ (LS)-γ (HPGe) Coincidence System

With the rapid development of the ADC, nowadays commercial acquisition card can sample the signal from the PMT directly with a sampling rate higher than 1GSPS. And signal from HPGe can also be sampled directly by commercial acquisition card with a 16bits sampling resolution which have an advantage in energy spectrum analysis. But the commercial acquisition system has following deficiencies: Firstly, signal generated by the PMT is a negative pulse, and signal generated by the HPGe is a positive pulse. Commercial acquisition system usually can't fit the input range of the signal or lose half of the ADC measurement range as most of the commercial acquisition cards are designed to sample the bipolar signal. Secondly, as the Fig.1 shown 4πβ (LS) −γ (HPGe) needs three high speed acquisition channel to sample the signal from PMT and two normal speed acquisition channel to sample the signal from HPGe respectively, so we should have all the acquisition channel synchronous with a same timestamp to do the 4πβ-γ coincidence[8]. It's hard to synchronize all the acquisition channels and even it's just impossible sometimes. Thirdly, to sample two different types of signals (β and γ signal), two kinds of acquisition cards should be used. But different acquisition cards usually have different working modes and output data formats. It's hard to unify all the acquisition channels. In addition, the max rate of the coincidence event is up to 10K per second, choosing a suitable method to record so much data becomes a difficult thing. Besides all the reasons above, the commercial acquisition card that can give consideration to all above-mentioned performance is quite expensive. That will be a large obstacle in generalizing the use of the system.

Herein, we introduce our home-made coincidence system based on FPGA and high speed ADC. All the hardware of this system is designed by ourselves for satisfying the requirement of the 4πβ-γ coincidence system fully. Three high speed acquisition channel can sample signal at a speed of 1GSPS to record the pre-amplified PMT signal. And two normal speed acquisition channel can sample signal with a 16bits resolution at a speed of 50MSPS. All the information about the decay signal such as amplitude and shape will be recorded with no prejudice. The logical module in FPGA can fulfill the function of MAC3. Different from commercial card, the acquisition channel of our system is synchronous within 500ps, which can decreases the workload of the off-line software and increase the coincidence accuracy at the same time. The data sampled by the acquisition channels will be processed by the FPGA firstly, and then stored in the DDR2 SDRAM for the PCI conflict detection. All the data in DDR2 will transmitted to host computer by PCI data bus and to be processed off line as Fig. 2 shown. In the test, we sampled the Si$^{90}$ LS signals and drawn the wave from the data sampled by the high-speed acquisition channel. We also sampled the γ signal of K$^{40}$ generated by the HPGe detector and drew the spectrum of the energy.

## 2  Data acquisition module

The basic structure of the coincidence is shown in Fig.2. In order to acquire the β signal, three acquisition cards with yellow color are utilized to sample the signal generated by 4πβ (LS) with three PMTs. Moreover, the β signal is only amplified by the fast amplifier. Compared with traditional 4πβ (LS), it can not only reduce the decay of the pulse signal but also decrease the conducted interference caused by other modules. The β signal from LS is a pulse with an amplitude of -5 ~ 0V and a time width from 4 to 20ns. To sample the signal directly, a sample rate up to 1GSPS is needed. As the pulse has a 5V margin, there should be a minimum voltage resolution of 0.1V for applying the acquisition module to the low-energy β nuclide as $H^3$. In order to interface with the FPGA with a low cost, the data bus to FPGA should not have a too high speed rate[9]. In fact, these are all achieved. TI ADC08D500 is chosen to support a sample rate up to 1GSPS by cross sampling with the ADC08D500's two ADC channels. The chip's data interface supports 1:2 DEMUX & LATCH technique, and it can turn the 8-bit data bus into 16-bit bus for each ADC channel[10]. Compared with other 1GSPS ADC, this chip in this system has an acquisition rate of 1GSPS at a data bus speed of 250M. With the chip's Double Data Rate (DDR) LVDS data interface support, the data interface's clock to FPGA is 125M, which can increase the stability of the FPGA. According to the final test, the ENOB of the 1GSPS channel @ 250MHz input is 7.2 (Signal generator: Tektronix AWG5000 series : 600MS/s maximum sampling rate, 2.5G bandwidth, 14 bits D/A resolution, 16M memory length). Although it is lower than 7.5 as shown in the datasheet of the device, it can still meet the requirement of the minimum voltage resolution.

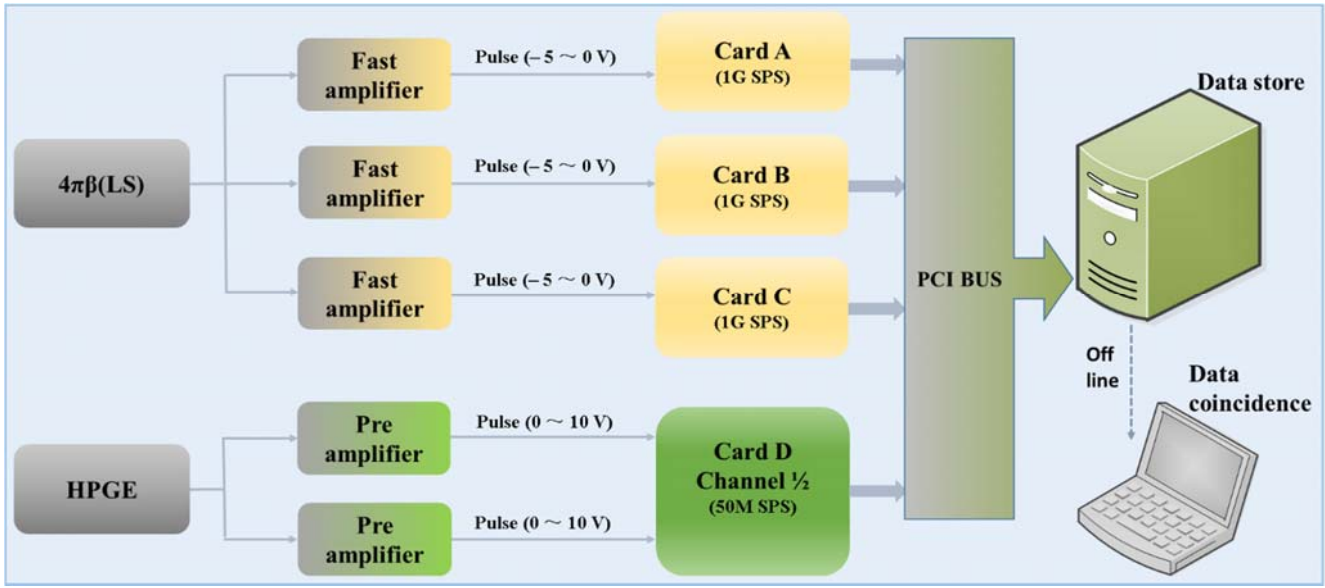

Fig. 2. Simplified block diagram of the 4πβ (LS)-γ (HPGe) digital coincidence system

For the acquisition of the two channels of γ signal from HPGe, a normal-speed acquisition card with two ADC channels of 16 bits and 50MSPS is utilized. As for the β signal, the pulse signal of the pre-amplifier is sampled. It difference from β signal is that the amplitude of γ signal is 0~10V and the time width of γ signal is above 1 us. As the γ signal is slower than the β signal, the acquisition rate of 50MSPS is enough to sample the signal. In order to analyze the spectra of the γ nuclide, the minimum voltage resolution is quite important. As a result, ADI AD9268 is chosen. The test on the acquisition channel shows that the SNR of the γ acquisition channel is 74.3 (signal generator ditto) at 10MHz, which is important in the spectra analysis of γ signal[11].

## 3 System channel synchronization

As shown in Fig. 3, three yellow high-speed acquisition cards and one green normal-speed acquisition card are plugged into the PXI case. In order to let all the acquisition channels sampling at the same timestamp, all the acquisition cards should operate under the condition of a same clock with a synchronous clock reset function. Fig. 4 shows the structure of the red card, which is utilized to synchronize the acquisition cards. The main function of this card is to distribute a reference clock of 50Mhz and send a synchronous reset signal called SYNC through the star-shaped trigger line on the PXI case. This card should be plugged into Slot 2 of the PXI case, as star-shaped trigger line is only provided at this location. Star-shaped trigger line is a trigger with a same time delay on PXI case, which is utilized to send a SYNC clock reset signal to enable all the timestamps reset on each acquisition card. Therefore, the timestamps have not only a same reset but also a same time counter. The synchronization module of each acquisition card utilizes a 48-bit timestamp, which would not be repeated within 65 days. The time is long enough to meet most measurement requirements.

The reason for distributing a clock of 50MHz is that, three acquisition cards with a sampling rate of 1G need a reference clock of 500MHz respectively, and the normal-speed acquisition card utilizes a reference clock of 50MHz. Consequently, the reference clock of 50MHz is the best choice to both increase time reliability and decrease the complexity of the synchronization module. In order to achieve the same delay time from each acquisition channel, cables with a same length are utilized. The clock of 50MHz is sent by the front panel as shown in Fig. 3.

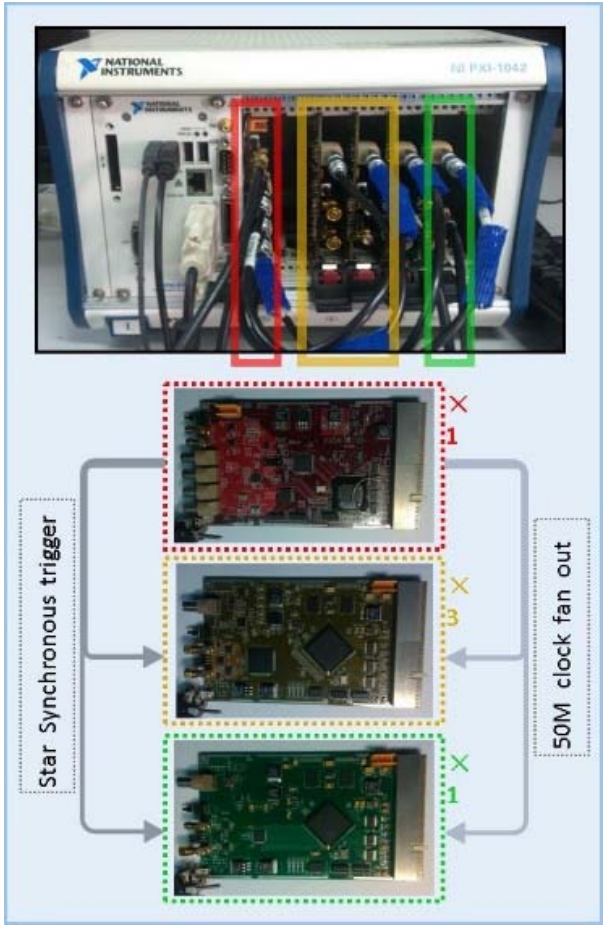

Fig. 3. PXI case with our coincidence system.

As the sampling interval of high-speed acquisition channel is 1ns, so the clock mismatch for all the high-speed acquisition channels be supposed to be within 500ps, and the normal-speed acquisition channel should also be synchronous to the high-speed channels. To meet this requirement, phase locked loop (PLL) chip AD9524 is applied in the synchronization module shown in Fig. 4. The AD9524 can not only control the phase of the clock to obtain a clock mismatch within 500ps, but also provide 4 fan-out low jitter LVDS clocks[12]. The synchronization between high-speed acquisition channels and normal-speed acquisition channels can be adjusted by setting the parameter of this chip. It also provides 4 LVDS reference clocks to each acquisition card. As worked out with the Oscilloscope (Agilent Infinii

Vision 7000B), the average jitters of the 4 LVDS clocks to each card are 12.08ps, 12.38ps, 12.10ps, and 12.44ps respectively. As the jitter will decrease the SNR of the ADC channels on the acquisition cards, each of the acquisition card shall also utilize an AD9524 to eliminate the jitter. For high-speed acquisition card, the PLL is also utilized to multiply the clock frequency to 500MHz.

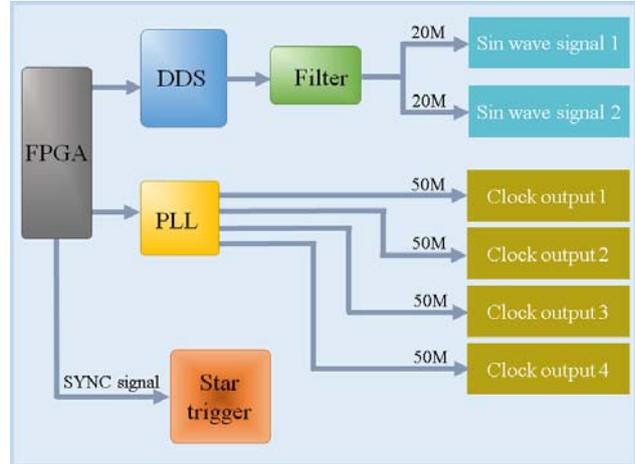

Fig. 4. Schematic diagram of the components of channels synchronization module.

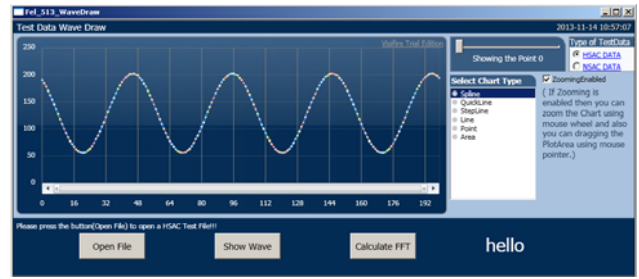

Fig. 5. The waveform sampled by the acquisition channels to calculate the mismatch of each channel.

To figure out the mismatch of each ADC channel, two 20M sine waves are fan-outed with a direct digital synthesizer (DDS). The sine waves are filtered by utilizing a band-pass filter to improve the signal to noise ratio (SNR). Each acquisition channel sample 4096 points from the sine waves at the same timestamp to calculate the phase, as shown in Fig. 5. Then, FFT function is adopted to calculate the phase.

A sine wave can describe as

$$F(\omega) = A(\omega) + jB(\omega)$$

The initial phase can be describe as

$$\theta = \tanh^{-1}(B(\omega)/A(\omega))$$

If the acquisition channel A have the initial phase θa, acquisition channel B is θb, then the phase difference between two acquisition channel is

$$\Delta\theta = \theta a - \theta b$$

After the adjustment of the PLL and the line. The average

$\Delta\theta$/rad of each acquisition channel is 0.003. Then the mismatch of each acquisition channel is

$$\Delta T = (\Delta\omega T_\omega)/(2\pi) = \frac{0.003}{2\pi} \times \frac{1}{20\times10^6} = 0.02387\text{ns}$$

The result reached to the requirement of the synchronization.

## 4 Logical module of FPGA

In order to reduce the complexity of the system design, the design scheme of the FPGA is the same, as shown in Fig. 6a. For high-speed acquisition cards, the interface for ADC is 32-bit DDR LVDS data bus with a clock of 125MHz. For normal-speed acquisition card, the interface for ADC bus is a 32-bit CMOS with a clock of 50MHz. The sampled data will be screened firstly to reduce the amount of the data. When a signal is larger than the pre-set trigger line, a trigger signal will be generated. A rejected time is also set to meet the special requirement of data screening. The screening module will enable the write function of the ADC data FIFO (first input first output) if all the requirements on the data stream are met. A data package will be recorded into the FIFO to wait to be sent to the DDR2 SDRAM. The data package, as shown in Fig. 6b, has 128-byte data or 256-byte data for selection. It can be set by the acquisition software to fit the length of the pulse. For most of the time, we use 128-byte data for β signal recording and 256-byte for γ signal recording.

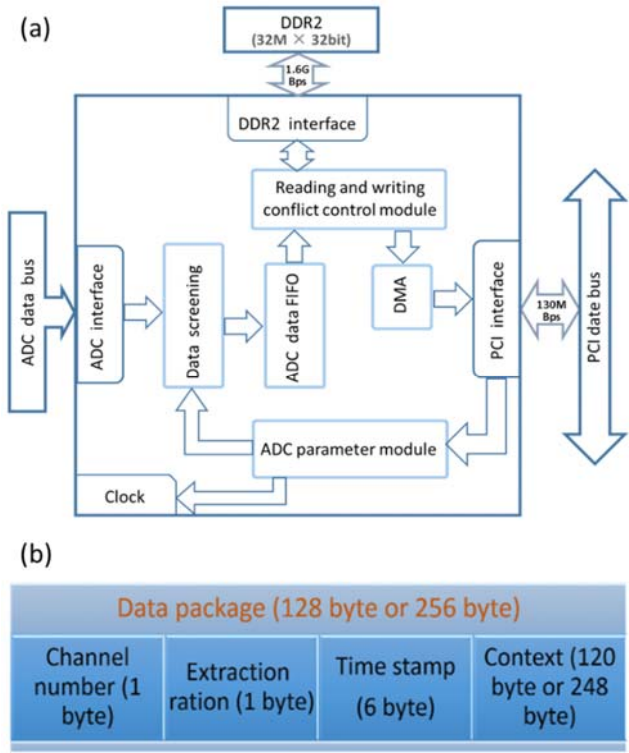

(a)

(b)

Data package (128 byte or 256 byte)

| Channel number (1 byte) | Extraction ration (1 byte) | Time stamp (6 byte) | Context (120 byte or 248 byte) |

Fig. 6. a) Schematic diagram of the structure of the FPGA; b) data package of the system.

All the data in DDR2 SDRAM would be transferred at a speed of 1M bytes per DMA burst. The PCI also transfers the configuration command to configure the PLL device and ADC. When there are 10K pulses per second, 8.75Mbyte of data should be transferred (10K×(3×128byte + 2×256)). The test on the system shows that the system is able to work very well at this pulse rate. According to the pressure test, it can be found that the system can also work well at a pulse rate of 20K, which can also meet the requirement of most measurements.

## 5 System testing

To test the coincidence system, firstly, we used the signal generator (Tektronix AWG5000 series) to simulate the pulse signal generated by the β and γ decay. As the AWG5000 only have two output channels, we used a channel to simulate the β decay and then fan-outed by the connector, same with the γ decay. The pulse of the β have a 20ns pulse width, the amplitude of the pulse is -0.5V. The pulse of the γ have a 2us pulse width, the amplitude of the pulse is 1V. The two channels were set to be external triggered by a random trigger source. The trigger rate is 20K/s. As the test, we found our system was working well, and all the pulses were recorded.

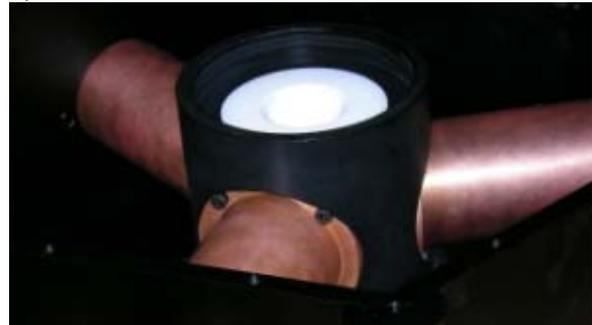

Fig. 7. The photo of 4πβ (LS) light chamber.

The real test environment was provided by the National Metrology Institute of China, the Fig.7 show the 4πβ (LS) light chamber. To compare with the traditional coincidence system, we used the Ultima Gold LLT as the liquid cocktail, $6.59\times10^3$Bq $H^3$ as the nuclide source. The signal from the PMTs was fan-outed to the traditional MAC3 and our system. Trigger line was set as -60mV. Test result shown as below:

| device | time(s) | counter rate | efficiency |
|---|---|---|---|
| MAC3 | 45 | 3455 | 52% |
| our system | 45 | 3455 | 52% |
| MAC3 | 4500 | 3447 | 52% |
| our system | 4500 | 3447 | 52% |

As the table shown above, it can fulfil the function of the traditional coincidence system.

To show the ability of the shape recording, we used the Ultima Gold AB as the liquid cocktail, $Si^{90}$ as the nuclide source, Fig. 8a show the shape of the pulse.

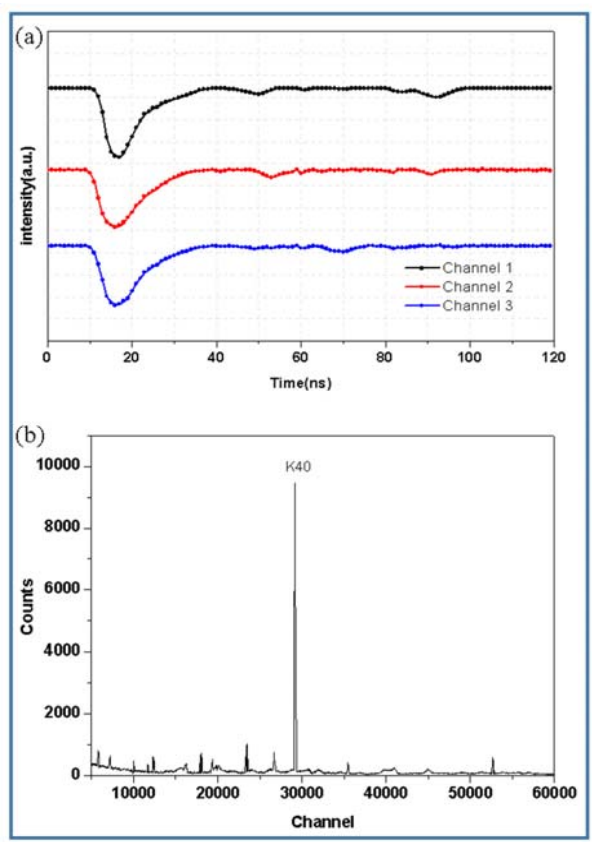

Fig. 8. a) The β signals from 4πβ (LS) in the measurement the Si$^{90}$; b) the energy spectrum of K$^{40}$.

Compared to Oscilloscope (Agilent Infinii Vision 7000B), we find it can well recovery the shape of the pulse.

For γ test, we used K$^{40}$ as our nuclide source, the Fig. 8b drawn the energy spectrum of K$^{40}$. The line around 30000 indicates the K$^{40}$. The small peaks of this spectrum are attributed to the other radionuclide contamination. Comparing to the traditional γ energy spectrum, our system have a higher resolution of the energy spectrum.

## 6 Conclusion

In this paper, a 4πβ (LSC)-γ (HPGe) coincidence system is present. The final test shown that, the hardware and the software we designed can work well to reach the requirement of 4πβ (LSC)-γ (HPGe). Further test will be done to find more method on radionuclide measurement on this coincidence system.

*Thanks to National Metrology Institute of China, they fund the project and provide us with a lot of help*

## References

1   Bobin C, Bouchard J, Pierre S et al, "Overview of a FPGA-based nuclear instrumentation dedicated to primary activity measurements," Appl. Radiat. Isot., 2012, 70: 2012

2   Kawadaa Y, Yamadaa T, Unnoa Y et al. A simple and versatile data acquisition system for software coincidence and pulse-height discriminationin 4πβ-γ coincidence experiments, Appl. Radiat. Isot., 2012, 70: 2031

3   Bobin C, Bouchard J, Thiam et al. Digital pulse processing and optimization of the front-end electronics for nuclear instrumentation, Appl. Radiat. Isot., 2014, 87: 193

4   Campion P J, The standardization of radioisotopes by the beta-gamma coincidence method using high efficiency detectors, Int. J. Appl. Radiat. Isot., 2014, 4: 193

5   Bobin C, Bouchard J, Censier B. First results in the development of an on-line digital counting platform dedicated to primary measurements, Appl. Radiat. Isot., 2010, 68: 1519

6   Bouchard J. MTR2: a discriminator and dead-time module used in counting systems, Appl. Radiat. Isot., 2000, 52: 44

7   Steele T, Mo L, Bignell L, Smith M et al. FASEA: a FPGA Acquisition System and Software Event Analysis for liquid scintillation counting, Nucl. Instrum. Methods A, 2009, 609: 217

8   Konorov I, Angerer H, Mann A et al. SODA: Time Distribution System for the PANDA Experiment, 2009 IEEE Nuclear Science Symposium Conference Record, Orlando, FL, USA, 2009.1863

9   Cyclone III Device Datasheet, Altera Corporation, July 2012.

10  ADC08D500 High Performance, Low Power, Dual 8-Bit, 500 MSPS A/D Converter, Texas Instruments Incorporated, May 2005

11  16-Bit, 80 MSPS/105 MSPS/125 MSPS, 1.8 V Dual Analog-to-Digital Converter (ADC), Analog Devices, D08123-0-9/09(A), 2009

12  Jitter Cleaner and Clock Generator with 6 Differential or 13 LVCMOS Outputs, Analog Devices, D09081-0-1/14(E), 2010